\documentclass[twocolumn]{aastex631}
\usepackage{amsmath}
\usepackage{amssymb}
\usepackage{hyperref}
\usepackage{soul}

\begin{document}

\title{Evidence for a Compact Stellar Merger Origin for GRB 230307A From Fermi-LAT and Multiwavelength Afterglow Observations}

\author[0000-0002-0170-0741]{Cui-Yuan Dai}
\affiliation{School of Astronomy and Space Science, Nanjing University, Nanjing 210023, China; hmzhang@nju.edu.cn; xywang@nju.edu.cn}
\affiliation{Key Laboratory of Modern Astronomy and Astrophysics (Nanjing University), Ministry of Education, Nanjing
210023, China}

\author[0000-0003-1698-0835]{Chen-Lei Guo}
\affiliation{School of Astronomy and Space Science, Nanjing University, Nanjing 210023, China; hmzhang@nju.edu.cn; xywang@nju.edu.cn}
\affiliation{Key Laboratory of Modern Astronomy and Astrophysics (Nanjing University), Ministry of Education, Nanjing
210023, China}

\author[0000-0001-6863-5369]{Hai-Ming Zhang}
\affiliation{School of Astronomy and Space Science, Nanjing University, Nanjing 210023, China; hmzhang@nju.edu.cn; xywang@nju.edu.cn}
\affiliation{Key Laboratory of Modern Astronomy and Astrophysics (Nanjing University), Ministry of Education, Nanjing
210023, China}

\author[0000-0003-1576-0961]{Ruo-Yu Liu}
\affiliation{School of Astronomy and Space Science, Nanjing University, Nanjing 210023, China; hmzhang@nju.edu.cn; xywang@nju.edu.cn}
\affiliation{Key Laboratory of Modern Astronomy and Astrophysics (Nanjing University), Ministry of Education, Nanjing
210023, China}

\author[0000-0002-5881-335X]{Xiang-Yu Wang}
\affiliation{School of Astronomy and Space Science, Nanjing University, Nanjing 210023, China; hmzhang@nju.edu.cn; xywang@nju.edu.cn}
\affiliation{Key Laboratory of Modern Astronomy and Astrophysics (Nanjing University), Ministry of Education, Nanjing
210023, China}

\begin{abstract}

GRB 230307A is the second-brightest gamma-ray burst (GRB) ever detected  over 50 yr of
observations and has a long duration in the prompt emission. Two galaxies are found to be close to the position of GRB 230307A: (1) a distant ($z \sim 3.87$) star-forming galaxy, located at an offset of $\sim 0.''2\operatorname{-}0.''3$ arcsec from the GRB position (with a projected distance of $\sim 1\operatorname{-}2 \, \rm kpc$); (2) a nearby  ($z= 0.065$) spiral galaxy, located at an offset of $30''$ (with a projected distance of $\sim 40 \, \rm kpc$).  Though it has been found that the brightest GRBs are readily detected in GeV emission by the Fermi Large Area Telescope, we find no GeV afterglow emission from GRB 230307A. Combining this with the optical and X-ray afterglow data, we find that a circumburst density as low as $\sim 10^{-5} \operatorname{-} 10^{-4}~{\rm cm^{-3}}$ is needed to explain the nondetection of GeV emission and the multiwavelength afterglow data, regardless of the redshift of this GRB. Such a low-density disfavors the association of GRB 230307A with the high-redshift star-forming galaxy, since the proximity of the GRB position to this galaxy would imply a higher-density environment. Instead,  the low-density medium is consistent with the circumgalactic medium, which agrees with the large offset between
GRB 230307A and the low-redshift galaxy. This points to the compact stellar merger origin for  GRB 230307A, consistent with the detection of an associated kilonova.

\end{abstract}

\section{Introduction} \label{sec:intro}


 Gamma-ray bursts (GRBs) can be classified based on their duration ($T_{90}$) into two categories: short GRBs with $T_{90} < 2 \, \mathrm{s}$ and long GRBs with $T_{90} > 2 \, \mathrm{s}$  \citep{Norris1984, Kouveliotou1993}.  Long
GRBs are commonly thought to  originate from the core collapse of massive stars
(e.g., \cite{Galama1998}) and short GRBs formed in the merger of two
compact objects (e.g., \cite{Abbott2017}).

While the duration of GRBs can generally provide insights into their origins, some GRBs defy classification based on duration alone.  Growing observations \citep{Gal2006, Gehrels2006, Ahumada2021, zhangbb2021} have shown that multiple criteria (such as supernova/kilonova associations and host galaxy properties) rather than burst duration only are needed to classify GRBs physically.  In this study, we propose that the circumburst density provides a supplement method to discern the origin of a particular GRB, GRB 230307A.

GRB 230307A was detected by the Fermi Gamma-ray Burst Monitor (GBM; \cite{Meegan2009}) at 15:44:06 UT on 2023 March 7, (denoted as $T_0$). The GBM light curve shows a burst with a duration ($T_{90}$) of approximately 35 s ($10-1000\, \rm keV$), and the fluence during $T_0-T_0+148\, \rm s$ is $(2.951 \pm 0.004) \times 10^{-3}\, \rm erg\,cm^{-2}$ \citep{GCN33441}, making it the second-brightest GRB  ever recorded over 50 yr of GRB
observations (after GRB 221009A) \citep{GBM2023}.  Two galaxies are found to be close to the position of GRB 230307A: (1) a distant ($z \sim 3.87$) star-forming galaxy \citep{2023Levan_z3}, located  $0.''3$ away from the GRB
position; (2) a nearby  ($z= 0.065$) spiral galaxy  \citep{2023Gillanders}),  located at an offset of $30''$.



 The high-redshift scenario fits well within the classification of GRB 230307A as a long burst from a young stellar population with a  chance coincidence probability of $P_{\text{cc}} \approx 0.03$ \citep{Yangyuhan2023}, which is reasonable for a physical association \citep{Bloom2002}, although it implies an unprecedented gamma-ray energy release ($10^{56}\, \text{erg}$). The low-redshift solution implies a more reasonable gamma-ray energy release ($3\times10^{53}\, \text{erg}$) and is consistent with a luminous red transient, which was found to be associated with this GRB, pointing to a kilonova at the redshift of $0.065$ \citep{Levan2023, Gillanders2023, Yangyuhan2023, 2023Sun}. However, the chance coincidence probability of this galaxy is $\sim 13\%$ \citep{Yangyuhan2023}, a value generally considered too high for a reliable physical association \citep{2013ApJ...776...18F, Tunnicliffe2014}, and GRBs with $P_{\text{cc}} > 0.1$ have been regarded as hostless in previous studies \citep{Connor2022}. Therefore, the host galaxy of GRB 230307A still needs to be studied carefully.


Brightest GRBs usually have GeV afterglow emission, as found in previous Large Area Telescope (LAT)-detected GRBs.  In the second catalog of LAT-detected GRBs, covering the first 10 yr of operations, a total of  169 GRBs are detected above 100 MeV. These GRBs confirm earlier findings that LAT primarily detects the brightest GBM bursts and
the high-energy emission is extended than the prompt emission. Motivated by this, we search for the GeV afterglow emission from GRB 230307A. The first observation of this GRB by Fermi/LAT
starts at 1570 s and ends at 7000 s after the GBM trigger.  To our surprise, no GeV afterglow emission is detected during this period and subsequent observations. In this paper, we will show that a circumburst density as low as $10^{-4}{\, \rm cm^{-3}}$ is needed to explain the nondetection of GeV emission and the multiwavelength afterglow emission, regardless of the redshift. Such a low-density medium disfavors the association with the distant ($z \sim 3.87$) star-forming galaxy considering the proximity between the GRB position and the host galaxy, but consistent with a circumgalactic medium far from the galaxy. The large offset between the GRB and the low-redshift  ($z= 0.065$)  galaxy fits well with the compact stellar merger origin for  GRB 230307A.

The paper is organized as follows: in \S\ref{sec_data}, we begin with analyzing Fermi-LAT observations of GRB 230307A and exploring the implications of Fermi-LAT nondetection. In \S\ref{sec_expl}, we show the method and results of afterglow modeling and constrain the environment density.  In \S\ref{sec_statistic_characteristic}, we point out that GRB 230307A is consistent with the statistical characteristics of short bursts rather than long bursts; and in \S\ref{sec_summary}, we provide a brief summary.

\section{Fermi-LAT Observation of GRB 230307A} 
\label{sec_data}

\subsection{Fermi-LAT Data Analysis}

We extract the Fermi-LAT extended type data for the GRB 230307A from the Fermi Science Support Center\footnote{\url{https://fermi.gsfc.nasa.gov}}. 
Only the photons $>100~\rm MeV$ within a $14\degr \times14\degr$ region of interest (ROI) centered on the Fermi-GBM position of GRB 230307A are considered for the analysis. 
The publicly available software \textit{fermitools} (v2.2.0) with the unbinned likelihood analysis method is performed in this work. 
The maximum zenith angle of 100$\degr$ is set to reduce the contamination from the gamma-ray Earth limb.
The instrument response functions (IRFs) (\textit{$P8R3\_SOURCE\_V3$}) are used.
The main background component consists of the isotropic emission template (``$iso\_P8R3\_SOURCE\_V3\_v1.txt$''), the diffuse Galactic interstellar emission template ($gll\_iem\_v07.fits$), and all sources listed in the fourth Fermi-LAT catalog  \citep{Abdollahi2020a}.
We select the observation time of GRB 230307A by LAT only when the angle from the Fermi-LAT boresight is less than $100\circ$. The first observation starts at 1570 s and ends at 7000 s after the GBM trigger.
Assuming a power-law spectrum of GRB 230307A, the test statistic (TS) value of the gamma-ray emission of this burst is found to be 9.2, implying that the significance of the gamma-ray emission for GRB 230307A is less than $5\sigma$. The upper-limit flux for GRB 230307A at a 95\% confidence level is shown in Table \ref{table_up}.

\subsection{Implication of Fermi-LAT Nondetection}
\label{sec_Fermi_non}

\begin{figure*}[htbp]
    \centering
    \includegraphics[width = 0.95\linewidth, angle = 0]{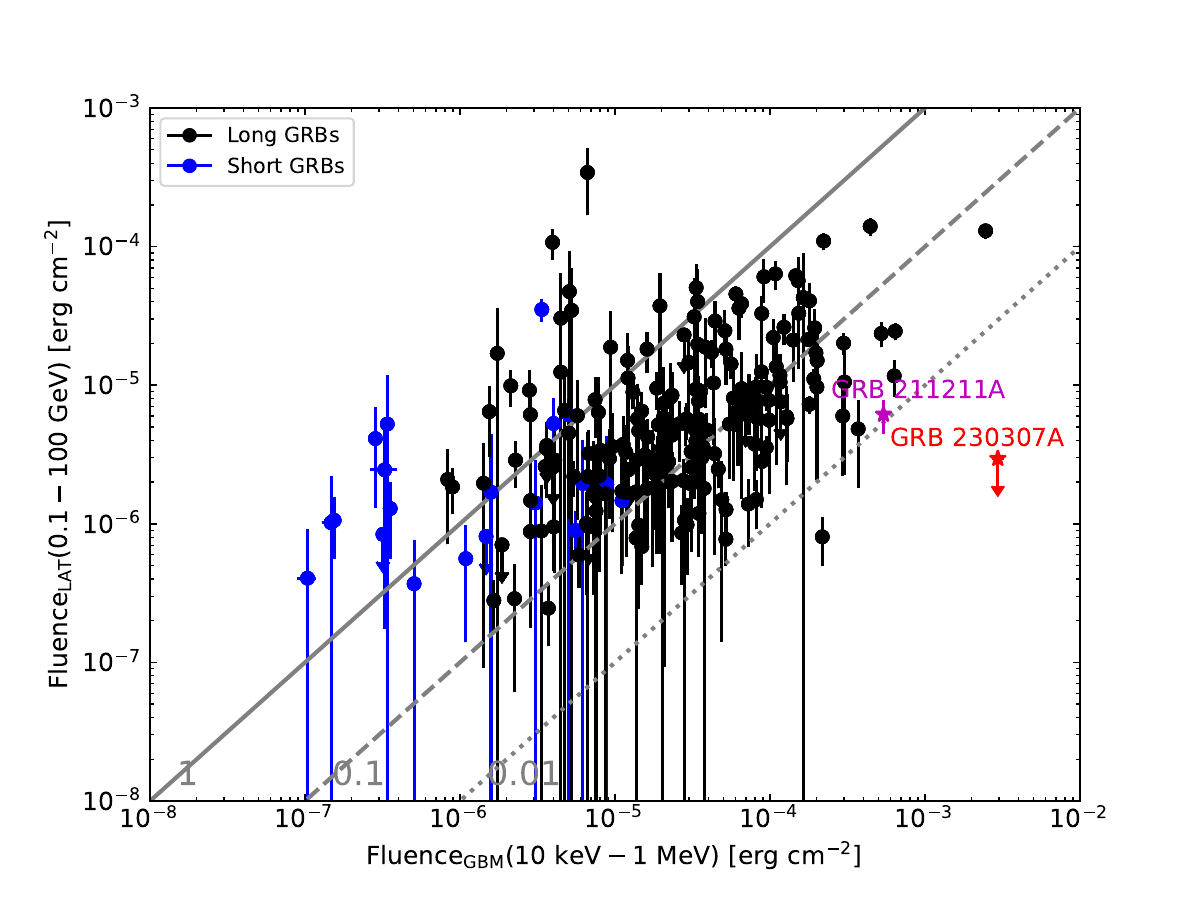} 
    \caption{A comparison between GRB 230307A and  GRBs in the Fermi-LAT Second Gamma-Ray Burst Catalog \cite{latgrb2019ApJ...878...52A} on the fluence  in 100 MeV-100 GeV measured by Fermi-LAT vs. the fluence in 10 keV-1 MeV measured by GBM. The black and blue data points represent long and short GRBs, respectively. The red star indicates the upper-limit measurement of GRB 230307A. {For comparison, we also show the measurement of GRB 211211A, denoted by the magenta star.} The solid, dashed, and dotted gray lines indicate where the ratio of the two fluences is 1, 0.1, and 0.01, respectively.}
    \label{LAT_grb}
\end{figure*}

{ The nondetection of GeV emission from such a bright GRB is surprising since it was found that LAT-detected GRBs have systematically higher fluences as compared to the whole GBM population (see Figure ~8 of  \cite{Nava2018IJMPD..2742003N} and Figure ~15 of  \cite{latgrb2019ApJ...878...52A}). In Figure ~\ref{LAT_grb}, we show the comparison of the measurement of LAT and GBM for GRB 230307A with other GRBs. 
The sample of the GRBs are obtained from Fermi-LAT Second Gamma-Ray Burst Catalog\footnote{LAT fluence is obtained with the highest TS; for details see \url{https://heasarc.gsfc.nasa.gov/W3Browse/fermi/fermilgrb.html}}\citep{latgrb2019ApJ...878...52A}.
One can see that the ratio between LAT and GBM fluences is smaller than 0.01 for GRB 230307A, in stark contrast to other short and long GRBs. We will show below that this anomaly is likely due to an unusually low-density environment of GRB 230307A.}

The afterglow synchrotron emission can extend to the energy window of Fermi-LAT ($\ga 100 {~\rm MeV}$). The synchrotron flux at the $h\nu_{\rm LAT}=200 {~\rm MeV}$  depends on the spectral regime of the observed frequency. 
In the case of $\nu_{\rm LAT} > \nu_{\rm c}$ ($\nu_{\rm c}$ is the characteristic synchrotron cooling frequency), the flux is
\begin{equation}
\label{Eq_Fv_LAT1}
F_\nu = 10^{(-2.86p+5.91)}\nu_{8}^{-\frac{p}{2}} \epsilon_B^{\frac{p-2}{4}} \xi_e^{2-p} \epsilon_e^{p-1} E_{53}^{\frac{p+2}{4}} t_{4}^{\frac{2-3p}{4}} D_{27}^{-2} \, \mu \rm Jy,
\end{equation}
where $h\nu_{8}$ is photon energy in units of
$100 \, \rm MeV$, $\epsilon_e$ and $\epsilon_B$ are the fractions of energy of the shocked gas in electrons and magnetic field,
$t$ is the time since the beginning of the explosion in the
observer frame (in units of  seconds), $E$ is the isotropic kinetic energy
in the forward shock,  $z$ is the redshift, and $D$
is the luminosity distance to the burst. We assume that only a fraction $\xi_e$ of shock-heated
electrons are accelerated into a power-law form.
In the case of $\nu_{\rm LAT} < \nu_{\rm c}$, the flux is
\begin{equation}
\begin{array}{ll}
\label{Eq_Fv_LAT2}
F_\nu = 10^{(-2.86p+10.9)}  {\nu_{8}}^{\frac{1-p}{2}}  \epsilon_B^{\frac{p+1}{4}} \xi_e^{2-p} \epsilon_e^{p-1} E_{53}^{\frac{p+3}{4}}n^{\frac{1}{2}} t_{4}^{\frac{3(1-p)}{4}}D_{27}^{-2} \,\mu \rm Jy.
\end{array}
\end{equation}
If $\nu_{\rm LAT} > \nu_{\rm c}$, as shown in Equation \ref{Eq_Fv_LAT1}, $F_\nu$ depends roughly on  $E/D^2$ (for $p\sim 2$), which is proportional to the gamma-ray fluence (for a given radiation efficiency) and the microphysical parameter $\epsilon_e$. Since the value of $\epsilon_e$ is not expected to differ substantially among different GRBs, one would not be able to explain the low GeV flux of GRB 230307A for this spectral regime.  Instead,  it is more natural to explain the low GeV flux in the spectral regime  $\nu_{\rm LAT} < \nu_{\rm c}$, since the flux depends on the density of the circumburst medium, which is expected to differ for different GRBs. According to Equation \ref{Eq_Fv_LAT2}, we have

\begin{equation}
    \begin{array}{ll} \label{Eq_n_LAT_up}
    n \leq 10^{5.72p-21.8} F_{\nu, 0}^2 {\nu_{8}}^{p-1}\epsilon_B^{\frac{-(p+1)}{2}} \\
    \times \xi_e^{2(p-2)} {\epsilon_e}^{2(1-p)} E_{53}^{\frac{-(p+3)}{2}} t_{4}^{\frac{3(p-1)}{2}} D_{27}^4 \, \rm cm^{-3} .
    \end{array}
\end{equation}

\begin{figure*}[htbp]

    \centering

    \includegraphics[width = 1\linewidth, angle = 0]{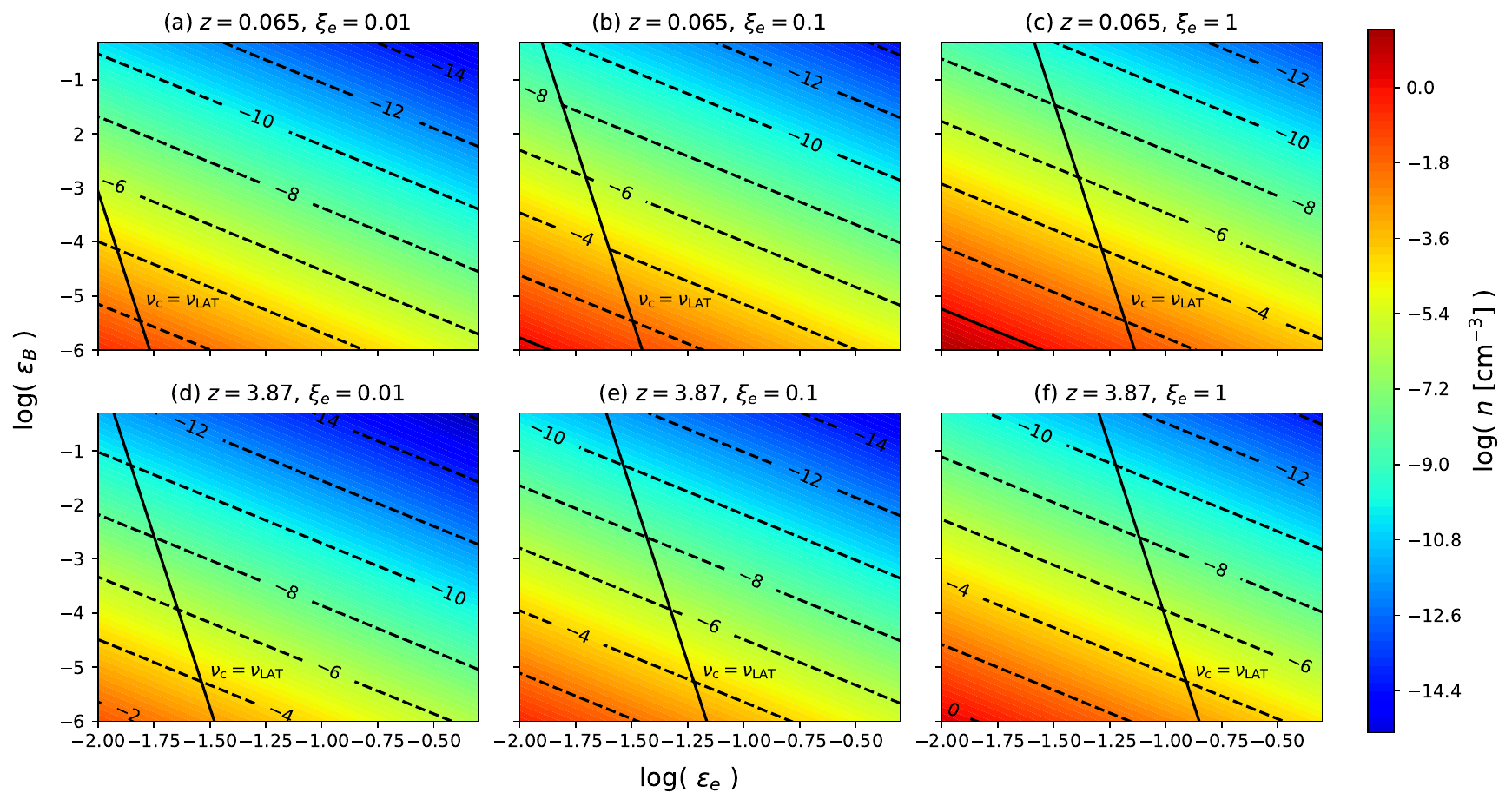}

    \caption{The parameter space diagram derived from the nondetection of  GeV afterglow emission. The contour of the number density is calculated using Equation \ref{Eq_n_LAT}, assuming that the radiation efficiency is $\eta_{\gamma} = 0.2$. The black solid line represents $\nu_{\rm LAT} = \nu_{\rm c}$. The area on the right of the black solid line satisfies $\nu_{\rm LAT} < \nu_{\rm c}$, indicating the allowed parameter space. Conversely, the region on the left of the black solid line is the excluded region. Panels (a) through (f) correspond to the following parameter value combinations: (a) $z = 0.065$, $\xi_e = 0.01$; (b) $z = 0.065$, $\xi_e = 0.1$; (c) $z = 0.065$, $\xi_e = 1$; (d) $z = 3.87$, $\xi_e = 0.01$; (e) $z = 3.87$, $\xi_e = 0.1$; (f) $z = 3.87$, $\xi_e = 1$.}
    \label{contour_LAT}
\end{figure*}

The observed X-ray photon index $\Gamma_{X}=1.73\pm0.10$ by XMM-Newton \citep{2023ApJ...956...97M} implies an electron spectrum index of $p=2.46\pm 0.2$ for $\nu_{X}<\nu_{\rm c}$. By employing the upper-limit flux provided by Fermi-LAT, as outlined in Table \ref{table_up}, and taking $p = 2.46$, one can derive an upper limit for the density of the circumburst medium:

\begin{equation} \label{Eq_n_LAT}
    n \leq \left \{
    \begin{array}{ll}
       10^{-2.5}\epsilon_{B, -4}^{-1.73} \xi_{e, -1}^{-2} k_{e, -1}^{-2.92} E_{53}^{-2.73} \, {\rm cm^{-3}} \,({\rm for}~z = 0.065), \\ \\
       10^{-5.0}\epsilon_{B, -4}^{-1.73} \xi_{e, -1}^{-2} k_{e, -1}^{-2.92} E_{57}^{-2.73} \, {\rm cm^{-3}} \,({\rm for}~z = 3.87),
   \end{array}
    \right.
\end{equation}
where  $k_e \equiv \epsilon_e/\xi_e$. Particle-in-cell (PIC) simulations find that the fraction of nonthermal electrons is about $\xi_{\rm e}\sim 1\% \operatorname{-} 10\%$, while    \cite{Sironi2013ApJ...771...54S,Duncan2023} constrain the fraction $\xi_e$ to a narrow distribution: $0.1 < \xi_e <1$ from the observations of the radio peaks in the light curves and spectral energy distributions in a sample of 49 radio afterglows. We thus consider the range of $0.01\leq\xi_e\leq 1$. 
We adjust the values of $\epsilon_e$ and $\epsilon_B$  to determine the parameter space for $n$ using Equation \ref{Eq_n_LAT} and upper limits in Table \ref{table_up}.  We consider three cases for $\xi_e$:    $\xi_e=0.01$,  $\xi_e=0.1$, and $\xi_e=1$. { We assume  a radiation efficiency of $\eta_{\gamma}=0.2$, so the isotropic kinetic energy is $E=(1-\eta_{\gamma})E_{\gamma, \rm iso}/\eta_{\gamma}=4E_{\gamma, \rm iso}$. }
The result is shown in Figure \ref{contour_LAT}  for both the low-redshift (upper panel) and high-redshift (bottom panel) cases. It is evident from the figure that GRB 230307A occurs in { a low-density medium ($n\le 10^{-3} \, {\rm cm^{-3}}$) under reasonable parameter spaces for $\epsilon_e$ and $\epsilon_B$ for both redshift cases. }

\section{Modeling  the Multiwavelength Afterglows of GRB 230307A } 
\label{sec_expl}

\subsection{Analytic Constraints on the Circumburst Density}
\label{sec_ana_cons_n}

A low-density medium is also inferred from X-ray and optical afterglows. As noted by \cite{Yangyuhan2023}, the X-ray afterglow of GRB 230307A is very dim compared with its MeV gamma-ray fluence. As the X-ray flux $F_{\nu_X}(1.12 \, \rm days) = 5.67 \times 10^{-2}\, \rm \mu Jy$  at $1 \, \rm keV$, the circumburst medium density could be derived as (assuming $p = 2.46$, see Equation \ref{Eq_Fv_LAT2})

\begin{equation} \label{eq_nn_x}
    n = \left \{
    \begin{array}{ll}
       10^{-3.8}\epsilon_{B, -4}^{-1.73} \xi_{e, -1}^{-2} k_{e, -1}^{-2.92} E_{53}^{-2.73} \, {\rm cm^{-3}} \,({\rm for}~z = 0.065), \\ \\
       10^{-6.3}\epsilon_{B, -4}^{-1.73} \xi_{e, -1}^{-2} k_{e, -1}^{-2.92} E_{57}^{-2.73} \, {\rm cm^{-3}} \,({\rm for}~z = 3.87).
   \end{array}
    \right.
\end{equation}

Further constraints on the density can be obtained by analyzing the peak in the optical light curve \citep{2023TESS}. The TESS observation of the optical ($I_{\rm c}$ band) afterglow of GRB 230307A shows a peak at $t_{I_{\rm c, p}} = 5.72 \times 10^{-2} \, \rm days$ after the GBM trigger. The peak in the optical afterglow light curve could have two origins: (1) the first scenario is the afterglow onset peak, corresponding to the deceleration time of the ejecta; (2) the second scenario for the peak is the crossing of the peak frequency (denoted as $\nu_{\rm m}$) over the observed frequency. The rising slope before the peak is different in the two scenarios. In the first scenario, the flux increases as $F_{\nu} \propto t^{3}$ for the spectral regime $\nu_{\rm m} < \nu < \nu_{\rm c}$ \citep{Yi2013}. In the second scenario, the flux $F_\nu$ grows with time $t$ following the relation $F_{\nu} \propto t^{1/2}$. The observed rising slope,  $0.6\pm 0.2$,  is consistent with the second scenario \citep{2023TESS}.  We also demonstrate that the peak time matches well with the crossing time of  $\nu_{\rm m}$ (see the Appendix.\ref{sec_proving_num_ism} for details). Denoting  $ t_{I_{\rm c,p}}$ as the crossing  time of $\nu_{\rm m}$ through the optical band, we obtain $F_{\nu, \rm max} = F_{\nu_{I_{\rm c}}}(t_{I_{\rm c}, \rm p})$ and $\nu_{\rm m} = \nu_{I_{\rm c}}$, where the flux density is $17.51\pm 0.07$ (TESS magnitude system) and $\nu_{I_{\rm c}}$ is the frequency corresponding to the pivot wavelength ($\lambda = 7839 \, \mathring{\rm A}$) of the TESS filter \citep{2023TESS}. Taking  $\nu_{\rm m} = \nu_{I_{\rm c}}$ at $t_{I_{\rm c, p}} = 5.72 \times 10^{-2} \, \rm days$, we  derive the value of $k_e$: $k_e = 0.55E_{53}^{-1/4}\epsilon_{B, -4}^{-1/4}$ and $k_e = 0.055E_{57}^{-1/4}\epsilon_{B, -4}^{-1/4}$ for the cases of $z = 0.065$ and $z = 3.87$, respectively (see Equation \ref{eq_num}). 
Then the density of the circumburst medium can be estimated as (see Equation \ref{eq_Fnu} and Equation \ref{eq_num})  
\begin{equation} \label{eq_nn_peak}
    n = \left \{
    \begin{array}{ll}
       10^{-5.7}\xi_{e, -1}^{-2} k_{e, -0.26}^{4} E_{53}^{-1} \, {\rm cm^{-3}} \,({\rm for}~z = 0.065), \\ \\
       10^{-5.2}\xi_{e, -1}^{-2} k_{e, -1.26}^{4} E_{57}^{-1} \, {\rm cm^{-3}} \,({\rm for}~z = 3.87),
   \end{array}
    \right.
\end{equation}

which also implies that GRB 230307A is situated in an exceptionally low-density environment for both redshifts.

\subsection{Multiwavelength Afterglow Modeling with  Afterglowpy}
\label{sec_fitting}

We perform a multiwavelength fit to the afterglow data of GRB 230307A using the public python package \emph{afterglowpy}.
\emph{Afterglowpy} is an open-source numerical and analytic modeling tool to calculate the synthetic light curve and spectra from an external shock \citep{Ryan2020}. 
To constrain the model parameter values and their errors, we use a Markov Chain Monte Carlo (MCMC) ensemble sampler with python package \emph{emcee} \citep{emcee2013PASP..125..306F} for the multiband fitting. 

When abundant multiwavelength afterglow data are available for some intensively studied GRBs, such as GRB 170817A and GRB 221009A, the jet is found to be not top-hat, but structured with some angular distribution in energy \citep{troja2018, Troja2017, Kasliwal2017}.  Simulations of jets initiated in realistic settings consistently result in more complex lateral energy profiles than a top-hat jet \citep{2021MNRAS.500.3511G}. Motivated by this, we consider Gaussian and power-law jet-type structures in the modeling of GRB 230307A. The angular energy distribution in Gaussian and power-law jets are given by \cite{Ryan2020}:
\begin{equation} \label{Eq_jet_model}
    \begin{array}{ll}
E(\theta) = E_0{\rm exp}\left( -\frac{\theta^2}{2\theta_{\rm c}^2} \right),  \, \rm Gaussian, \\ 
E(\theta) = E_0\left(1 + \frac{\theta^2}{b\theta_{\rm c}^2} \right)^{-b/2},  \, \rm Power\operatorname{-}law.
   \end{array}
\end{equation}
Each  model is parameterized by a normalization $E_0$,
a width $\theta_{\rm c}$, and a truncation angle $\theta_{\rm w}$ outside of which the
energy is initially zero \citep{Ryan2020}. The power-law model has an additional
power-law index $b$, which describes the steepness of the power-law
profile.

We adopt a log-uniform prior distribution for several parameters, including the isotropic kinetic energy $E$, the half-width of the jet core $\theta_{\rm c}$, the truncation angle $\theta_{\rm w}$, the number density of the circumburst medium $n$, the electron participation fraction $\xi_e$, the equipartition factor in electrons $\epsilon_e$ and in magnetic field $\epsilon_B$. Additionally, a uniform prior distribution is applied to the electron distribution power-law index $p$ and the power-law index $b$ in the power-law jet model. Based on the extreme brightness of this GRB, we assume that the jet is on-axis, i.e., the viewing angle is $\theta_v=0$  in the modeling. Initially, we fit the parameters using all the observational data and find that the JWST observations at approximately 60 days exceed the predicted light curve, which is consistent with the result in \cite{2023JWST}. This indicates the presence of an additional infrared source besides the afterglow of GRB 230307A. Therefore, we regard the optical flux observed by JWST around 60 days merely as upper limits for the afterglow emission.

\begin{figure*}[htbp]
\centering
    \includegraphics[width = 0.49\linewidth]{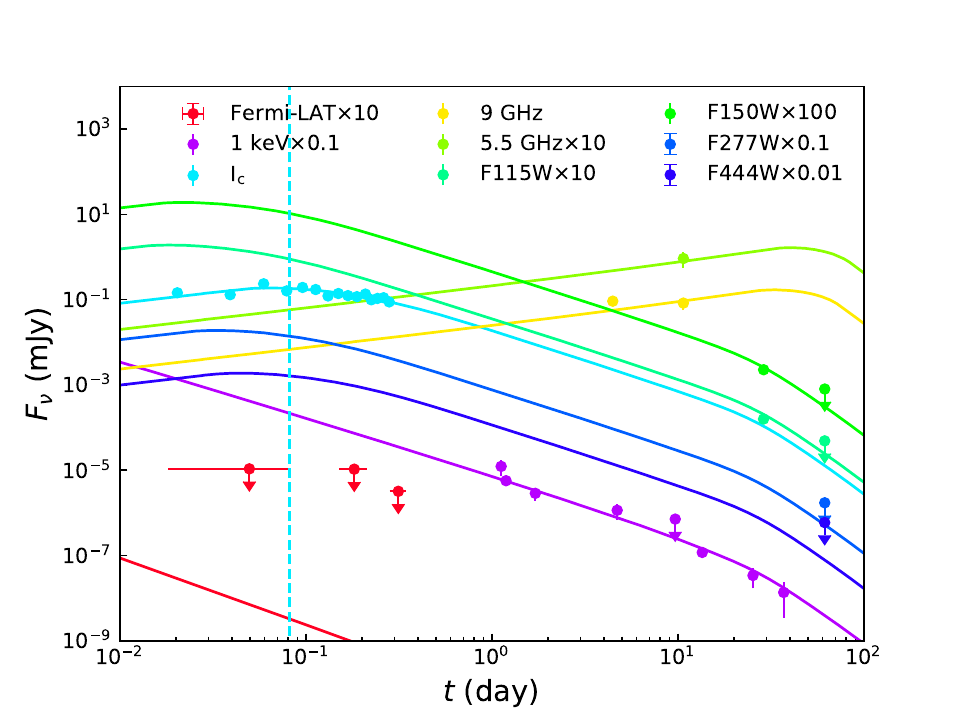}
    \includegraphics[width = 0.49\linewidth]{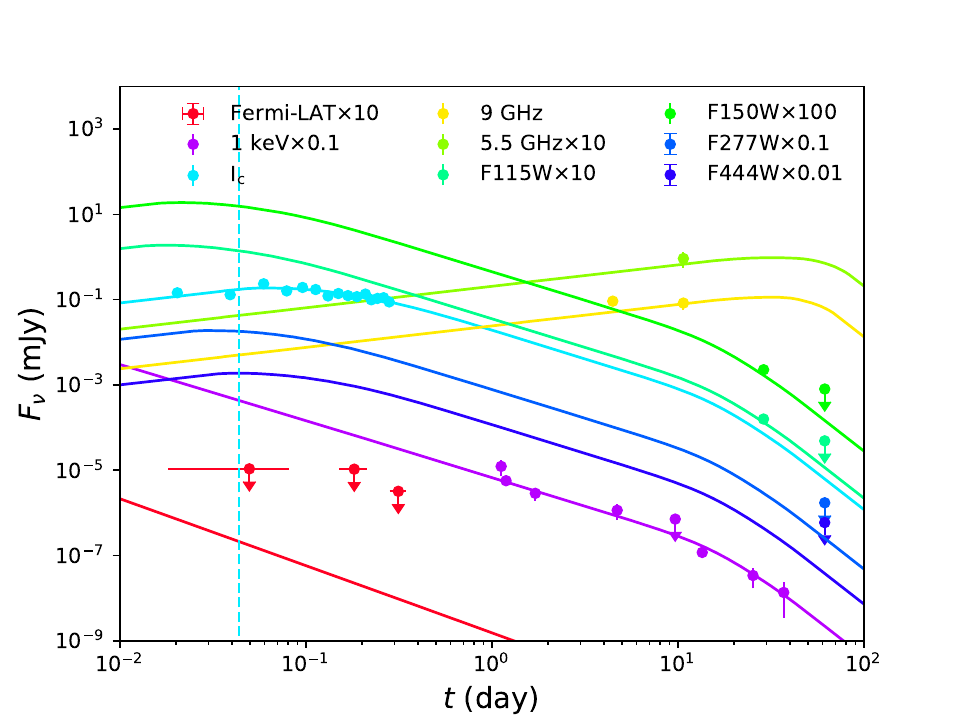}
    \caption{The best-fit model for the multiwavelength light curve of GRB 230307A  using the Gaussian jet model  in the cases of $z = 0.065$ (left panel) and $z = 3.87$ (right panel). The cyan vertical dashed lines mark the time when $\nu_{\rm m}=\nu_{I_{\rm c}}$, determined through the analytic method \cite{Sari1998}.}
    \label{fig_bestfit}
\end{figure*}

\begin{figure*}[htbp]
    \includegraphics[width = 1\linewidth, angle = 0]{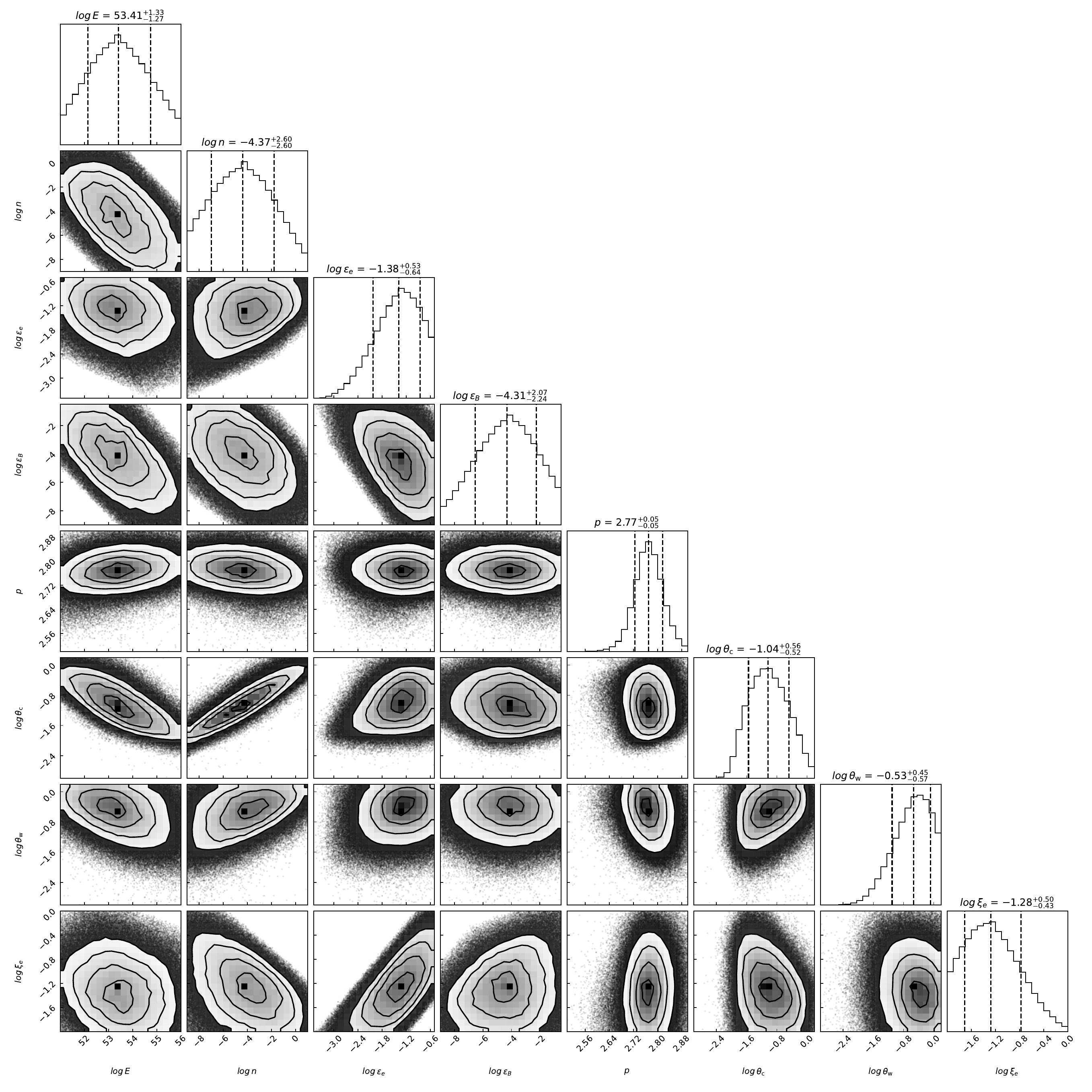}  
    \caption{Corner plot demonstrating the properties of the 8-dimensional posterior obtained from our MCMC sampling with the Gaussian jet model in the case of $z = 0.065$. The meaning of the parameters is explained in the text. Dashed lines and labels along the diagonal indicate the median value and symmetric $68\%$ uncertainties (corresponding to the 16 and 84 quantiles) for each parameter's marginalized distribution. The results of best-fit parameters in other cases can be found in Table \ref{best_fit_parameter}.}
    \label{fig_emcee}
\end{figure*}

\begin{figure*}[htbp]
    \includegraphics[width = 1\linewidth, angle = 0]{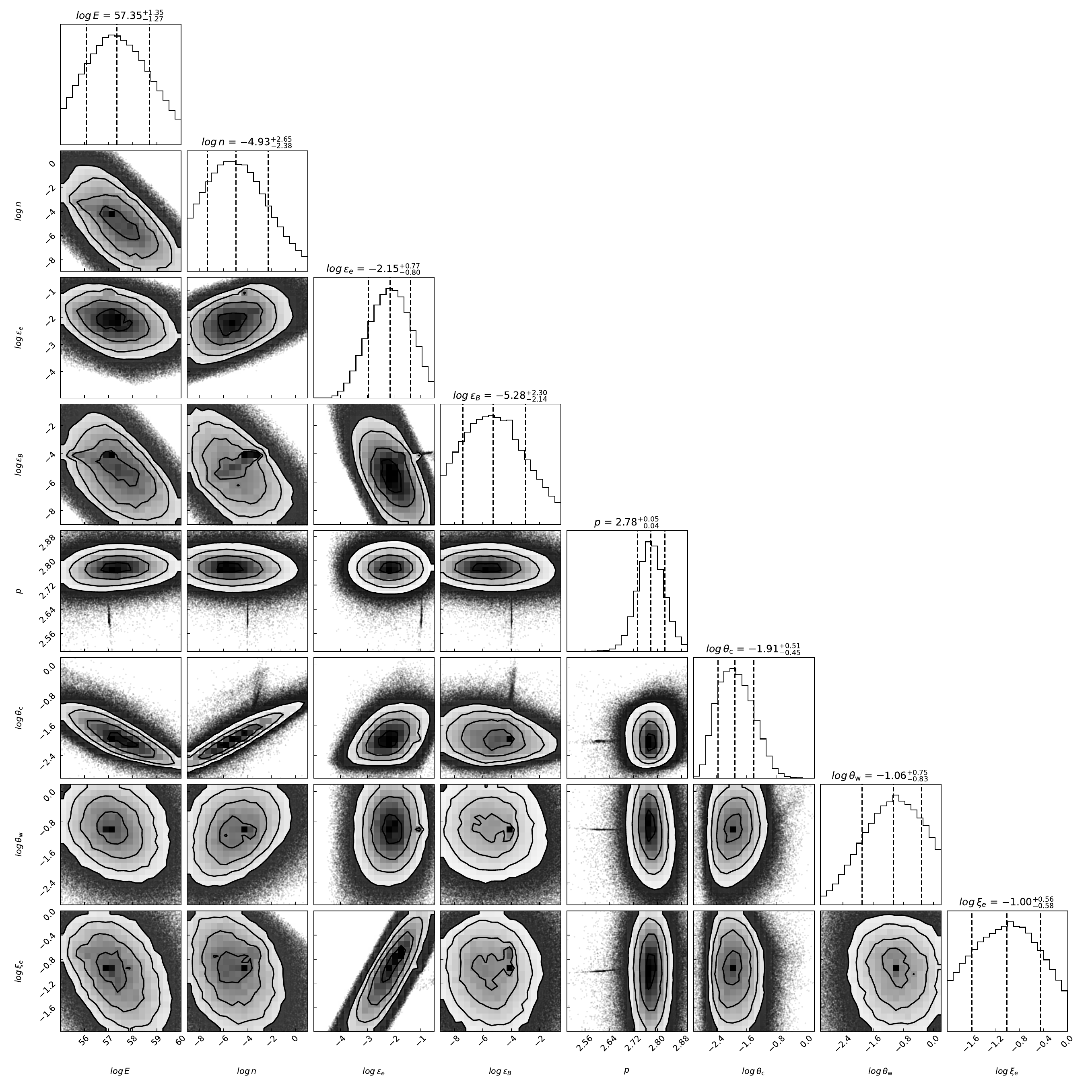}
    \caption{Same as Figure \ref{fig_emcee}, but for the case of  $z=3.87$.}
    \label{fig_emcee_387}
\end{figure*}

The values of the fitting parameters for the Gaussian and power-law jet-type models are depicted in Table \ref{best_fit_parameter}. The best-fit light curves, determined by minimizing the reduced$\operatorname{-}\chi^2$ across all completed MCMC runs, are presented in Figure \ref{fig_bestfit} for both redshift scenarios in the Gaussian jet model. The parameter distribution for the Gaussian jet model is shown in Figures \ref{fig_emcee} and \ref{fig_emcee_387} for the case of $z = 0.065$ and $z = 3.87$, respectively. The density of the circumburst medium is low in both redshift cases, with $n$ approximately in the range of $10^{-5} \operatorname{-} 10^{-4} \, \rm cm^{-3}$, which is consistent with the analytical constraints derived from Fermi-LAT and multiband observations (see Figure \ref{contour_LAT}, Equation \ref{eq_nn_x} and Equation \ref{eq_nn_peak}).  For such a low-density medium, GRB is expected to originate from a merger event rather than the collapse of a massive star, as the latter would typically occur in a star-forming region with a higher density. In other words, if GRB 230307A is situated in a star-forming galaxy with $z = 3.87$, the small offset of $0.''3$ \citep{Yangyuhan2023, 2023JWST} (with a projected distance of $\sim 1\operatorname{-}2 \, \rm kpc$) would be inconsistent with such a low density.   Instead, this low-density medium is consistent with the circumgalactic medium \citep{2017ARA&A..55..389T}, which agrees with the large offset between  GRB 230307A and the low-redshift  ($z= 0.065$) galaxy.  The large offset agrees with the spatial distribution of short GRBs \citep{2013ApJ...776...18F, Fong2022}, pointing to a compact stellar merger origin for the long-duration GRB 230307A (see \S\ref{sec_statistic_characteristic}). Thus, our result provides independent evidence for compact stellar merger origin for  GRB 230307A, besides the evidence from the detection of an associated kilonova \citep{2022Rastinejad,2022Natur.612..228T,2022Natur.612..236M}.

\section{Statistic Characteristics}
\label{sec_statistic_characteristic}

In this section, we suggest that, except for the duration \(T_{90}\), GRB 230307A agrees well with the statistical characteristics (offset from the galaxy center to the burst and circumburst density) of short bursts rather than long bursts, thereby providing further support for the merger origin of GRB 230307A.

\begin{figure*}[htbp]
    \includegraphics[width = 1\linewidth, angle = 0]{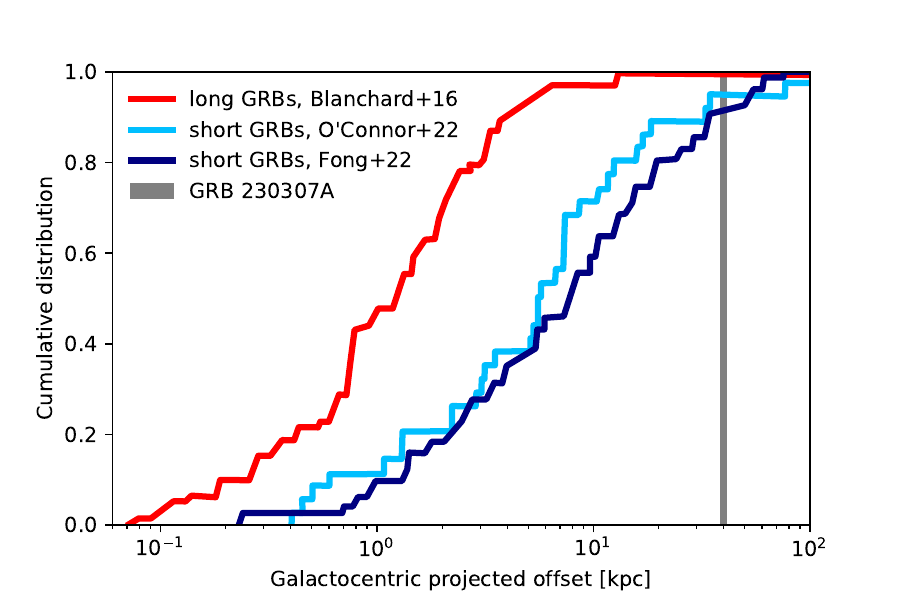}
    \caption{The distribution of galactocentric projected offsets for GRBs. The samples for short GRBs are extracted from \citep{Connor2022, Fong2022}, whereas the sample for long GRBs is obtained from \citep{Blanchard2016}.}
    \label{fig_distri}
\end{figure*}

Several statistical analyses of the offset from the galaxy center to the burst have been conducted in previous studies \citep{Connor2022, 2013ApJ...776...18F, Fong2022, Blanchard2016}. According to the  GRB offset sample in \cite{Connor2022} and \cite{Fong2022}, it is evident that the offset of GRB 230307A is more consistent with the short burst sample  (see Figure  \ref{fig_distri}):  approximately \(5\%-10\%\)  of short bursts exhibit offsets larger than \(40 \, \rm kpc\) and the maximum offset can extend to approximately \(\sim 60 \, \rm kpc\). 
In contrast, the offset in long bursts is limited to \(\sim 10 \, \rm kpc\), and the large offset (\(\sim 40 \, \rm kpc\)) of GRB 230307A is inconsistent with the long burst sample.

Additionally, the circumburst density of GRB 230307A is also more consistent with that of short bursts. \cite{Connor2020} constrained the density of short bursts using a short GRB sample and found that \(16\%\) of bursts have circumburst density of \(\lesssim 10^{-4} \, \rm cm^{-3}\). On the other hand, it is well known that long GRBs are concentrated in the brightest regions of their host galaxies \citep{Blanchard2016, Fruchter2006} and the circumburst density for long GRBs  are found to be  inconsistent with a low density of   \(n \lesssim 10^{-4} \, \rm cm^{-3}\) \citep{2001Panaitescu, 2003Bloom}.


Therefore, we conclude that GRB 230307A more closely resembles a short burst based on the statistical characteristics (offset and circumburst density) of  GRBs.

\section{Summary}\label{sec_summary}
GRB 230307A is the second-brightest GRB ever recorded  over 50 yr of GRB
observations, but its afterglow emission is very dim. We find no GeV emission from such a bright GRB through analyzing the Fermi-LAT observations,  which can be explained if this GRB occurs in a low-density medium. Modeling of the multiwavelength afterglow data also constrains the density to be as low as $10^{-4}{\rm cm^{-3}}$, regardless of the redshift of this GRB. Such a low density disfavors the association of GRB 230307A with the star-forming galaxy at $z \sim 3.87$, because the proximity of the GRB position to this galaxy would imply a higher density, as normally expected from massive star-forming regions. Instead, this low-density medium is consistent with the circumgalactic medium, which agrees with the large offset between  GRB 230307A and the low-redshift  ($z= 0.065$) galaxy. This also strongly supports the compact stellar merger origin for the long-duration GRB 230307A, consistent with the detection of an associated kilonova. 

GRB 230307A exhibits similar characteristics to GRB 211211A, which is also a long-duration burst with a compact stellar merger origin. These two bursts have comparable levels of afterglow flux intensity in both optical and X-ray bands. Interestingly, GRB 211211A is also found to occur in a low-density medium with $n \sim 10^{-4} \, \rm cm^{-3}$  \citep{zhang2022}, implying that the two peculiar bursts occur in quite similar environments.

\begin{acknowledgments}
The work is supported by the NSFC under grants Nos. 12333006, 12121003, 12203022, and U2031105,  and the Natural Science Foundation of Jiangsu Province grant BK20220757.

\end{acknowledgments}

\begin{table*}
\centering
\renewcommand\arraystretch{0.8}
\caption{{The  upper limits (at the 95\% confidence level)  on the GeV afterglow flux derived from Fermi-LAT data analysis.}}
\label{table_up}
\begin{tabular}{ccc}

\hline\hline
Time Interval since $T_0$ & Flux (0.1--10 GeV) & Flux Density (at 200 MeV) 
\\

(s) & $10^{-9} \, \rm (erg/cm^2/s)$ & $10^{-6} \, \rm (mJy)$
\\
\hline
1570--7000 & $< 1.80$ & $< 1.08$
\\
13000--18400 & $< 1.77$ & $< 1.06$
\\
24360--29830  & $< 0.54$ & $< 0.32$
\\
\hline\hline
\end{tabular}
\end{table*}

\begin{table*}
\centering
\caption{The best-fit parameters for the structured jet models in the two redshift cases, with uncertainties provided at the $1 \, \sigma$ Confidence Levels. The reduced$\operatorname{-} \chi^2$ represents the minimum value obtained overall completed runs and dof represents the degrees of freedom.}
\label{best_fit_parameter}
\begin{tabular}{lccc}
\hline\hline
Parameter & Prior & Posterior (Gaussian jet) & Posterior (Power-law jet)
\\
\hline
z=0.065
\\
$log\,E \,(\rm erg)$ & $[51, 55]$ & $53.41^{+1.33}_{-1.27}$ & $53.33^{+1.33}_{-1.22}$ 
\\
$log\, n \,(\rm cm^{-3})$ & $[-9, 1]$ & $-4.37\pm 2.60$ & $-4.29^{+2.60}_{-2.57}$ 
\\
$log\, \epsilon_e$ & $[-5, -0.5]$ & $-1.38^{+0.53}_{-0.64}$ & $-1.37^{+0.51}_{-0.63}$ 
\\
$log\, \epsilon_B$ & $[-9, -0.5]$ & $-4.31^{+2.07}_{-2.24}$ & $-4.26^{+2.03}_{-2.23}$ 
\\
$p$ & $[2.01, 2.9]$ & $2.77\pm 0.05$ & $2.77\pm 0.05$ 
\\
$log\, \theta_{\rm c}\, (\rm rad)$ & $[-3, log(\pi/2)]$ & $-1.04^{+0.56}_{-0.52}$ & $-1.01^{+0.56}_{-0.54}$ 
\\
$log\, \theta_{\rm w}\, (\rm rad)$ & $[-3, log(\pi/2)]$ & $-0.53^{+0.45}_{-0.57}$ & $-0.54^{+0.44}_{-0.58}$ 
\\
$log\,\xi_e$ & $[-2, 0]$ & $-1.28^{+0.50}_{-0.43}$ & $-1.27^{+0.50}_{-0.43}$
\\
$b$ & $[0.1, 10]$ & $\operatorname{-}$ & $5.09^{+2.87}_{-2.93}$ 
\\
$\chi^2/\rm dof$ & $\operatorname{-}$ & $1.84$ & $1.92$
\\
\hline
z=3.87
\\
$log\,E \, \rm (erg)$ & $[55, 60]$ & $57.33^{+1.35}_{-1.27}$ & $57.36^{+1.34}_{-1.21}$
\\
$log\, n \, \rm (cm^{-3})$ & $[-9, 1]$ & $-4.93^{+2.65}_{-2.38}$ & $-5.11^{+2.79}_{-2.17}$
\\
$log\, \epsilon_e$ & $[-5, -0.5]$ & $-2.15^{+0.77}_{-0.80}$ & $-2.15^{+0.77}_{-0.80}$
\\
$log\, \epsilon_B$ & $[-9, -0.5]$ & $-5.28^{+2.30}_{-2.14}$ & $-5.27^{+2.22}_{-2.13}$
\\
$p$ & $[2.01, 2.9]$ & $2.78^{+0.05}_{-0.04}$ & $2.78^{+0.06}_{-0.05}$
\\
$log\, \theta_{\rm c} \, \rm (rad)$ & $[-3, log(\pi/2)]$ & $-1.91^{+0.51}_{-0.45}$ & $-1.88^{+0.62}_{-0.50}$
\\
$log\, \theta_{\rm w} \, \rm (rad)$ & $[-3, log(\pi/2)]$ & $-1.06^{+0.75}_{-0.83}$ & $-1.05^{+0.73}_{-0.85}$
\\
$log\,\xi_e$ & $[-2, 0]$ & $-1.01^{+0.56}_{-0.58}$ & $-1.00^{+0.55}_{-0.58}$ 
\\
$b$ & $[0.1, 10]$ & $\operatorname{-}$ & $5.34^{+2.66}_{-2.98}$
\\
$\chi^2/\rm dof$ & $\operatorname{-}$ & $1.91$ & $2.01$
\\
\hline\hline
\end{tabular}
\end{table*}

\clearpage
\appendix 
\setcounter{equation}{0}
\renewcommand\theequation{A\arabic{equation}}

\section{The evidence of peak caused by $\nu_{\rm m}$ crossing the optical band}
\label{sec_proving_num_ism}
According to  \cite{Sari1998}, the synchrotron spectrum of a relativistic shock with a power-law electron distribution can be given as follows:
\begin{equation} \label{eq_spectrum}
    F_\nu = \left \{
    \begin{array}{ll}
       (\nu/\nu_{\rm c})^{1/3}F_{\nu, \rm{max}}, &\nu_{\rm m} > \nu_{\rm c} > \nu, \\
       (\nu/\nu_{\rm c})^{-1/2}F_{\nu, \rm{max}}, &\nu_{\rm m} > \nu > \nu_{\rm c}, \\
       (\nu_{\rm m}/\nu_{\rm c})^{-1/2}(\nu/\nu_{\rm m})^{-p/2}F_{\nu, \rm{max}}, &\nu > \nu_{\rm m} > \nu_{\rm c}, \\
       (\nu/\nu_{\rm m})^{1/3}F_{\nu, \rm{max}}, &\nu_{\rm c} > \nu_{\rm m} > \nu, \\
        (\nu/\nu_{\rm m})^{-(p-1)/2}F_{\nu, \rm{max}}, &\nu_{\rm c} > \nu > \nu_{\rm m}, \\
       (\nu_{\rm c}/\nu_{\rm m})^{-(p-1)/2}(\nu/\nu_{\rm c})^{-p/2}F_{\nu, \rm{max}}, &\nu > \nu_{\rm c} > \nu_{\rm m},
   \end{array}
    \right.
\end{equation}
where the observed peak flux at distance D from the source $F_{\nu,\rm max}$ and the break frequencies $\nu_{\rm m}$ and $\nu_{\rm c}$ can be written as  \citep{Sari1998}:
\begin{align}
    F_{\nu ,\rm{max}} &= 1.1\times 10^5\xi_e\epsilon_B^{1/2}E_{52}n^{1/2}D_{28}^{-2}\, \mu \rm{Jy}, \label{eq_Fnu} \\
     \nu_{\rm m} &= 5.7\times 10^{14}\epsilon_B^{1/2}\xi_e^{-2}\epsilon_e^2E_{52}^{1/2}t_d^{-3/2}\, \rm{Hz}, \label{eq_num} \\
    \nu_{\rm c} &= 2.7\times 10^{12}\epsilon_B^{-3/2}E_{52}^{-1/2}n^{-1}t_d^{-1/2}\, \rm{Hz} \label{eq_nuc}.
\end{align}

From Equation \ref{eq_num}, we can infer that  $\nu_{\rm m}(t) = \nu_{\rm m}(t_{I_{\rm c, p}})(t/t_{I_{\rm c}, \rm p})^{-3/2}$. According to Equation \ref{eq_spectrum}, the flux of radio band is given by ($\nu < \nu_m < \nu_{\rm c}$)

\begin{align} \label{eq_expected_flux_r}
    F_{\nu}(t) = \left[ \frac{\nu}{\nu_{\rm m}(t_{I_{\rm c, p}})} \right]^{1/3}\left( \frac{t}{t_{I_{\rm c, p}}} \right)^{1/2}F_{\nu, \rm max},
\end{align}

The photon index $\Gamma_{X}=1.73\pm0.10$ of the X-ray afterglow observed by XMM-Newton  \citep{2023ApJ...956...97M} suggests that  $\nu_{\rm m} < \nu_{\rm X} < \nu_{\rm c}$ at several days after the burst, so the afterglow is in slow-cooling regime, i.e.,  $\nu_{\rm m} < \nu_{\rm c}$. The flux of X-ray band ($\nu_m < \nu < \nu_{\rm c}$) is
\begin{align} \label{eq_expected_flux_x}
    F_{\nu}(t) = \left[ \frac{\nu}{\nu_{\rm m}(t_{I_{\rm c, p}})} \left( \frac{t}{t_{I_{\rm c, p}}} \right)^{3/2} \right]^{-(p-1)/2}F_{\nu, \rm max}.
\end{align}

Combing Equation \ref{eq_expected_flux_r} and Equation \ref{eq_expected_flux_x}, we can derive that

\begin{equation} \label{vm}
    \nu_{\rm m}(t_{I_{\rm c, p}}) = \left[ \nu_{r}^{-\frac{1}{3}}{\nu_{X}}^{\frac{1-p}{2}}\frac{F_{\nu_{r}}(t_{r})}{F_{\nu_ {X}}(t_{X})} \left( {\frac{t_{r}}{t_{I_{\rm c, p}}}} \right)^{-\frac{1}{2}} \left( {\frac{t_{X}}{t_{I_{\rm c, p}}}} \right)^{\frac{3(1-p)}{4}} \right]^{\frac{6}{1-3p}},
\end{equation}

Once we know how the radio flux ($F_{\nu_{r}}$) and X-ray flux ($F_{\nu_ {X}}$) vary with time and the electron spectrum index $p$, we can calculate the break frequency $\nu_{\rm m}$ at $t_{I_{\rm c, p}}$.

According to Equation \ref{eq_num}, the time $\nu_{\rm m}$ crossed by $\nu_{I_{\rm c}}$ is

\begin{equation}\label{eq_tm}
    t_{\rm m}(\nu_{I_{\rm c}}) = t_{I_{\rm c, p}} \left[\frac{\nu_{I_{\rm c}}}{\nu_{\rm m}(t_{I_{\rm c, p}})} \right]^{-2/3}.
\end{equation}


Since $t_{I_{\rm c, p}} = {\rm max} \{t_{\rm dec}, t_{\rm m}(\nu_{I_{\rm c}}) \}$, which implies $\nu_{\rm m} < \nu_{I_{\rm c}} < \nu_{\rm c}$ after $t_{I_{\rm c, p}}$, the flux in optical band can be written as

\begin{align} \label{eq_expected_flux_o}
    F_{\nu}(t) = \left[ \frac{\nu}{\nu_{\rm m}(t_{I_{\rm c. p}})} \left( \frac{t}{t_{I_{\rm c, p}}} \right)^{3/2} \right]^{-(p-1)/2}F_{\nu, \rm max}.
\end{align}

Combing Equation \ref{eq_expected_flux_x} and Equation \ref{eq_expected_flux_o}, the electron spectrum index $p$ is given by

\begin{align}\label{eq_p}
    p &= 1-2\frac{ln(F_{\nu_{X}}(t_{X})/F_{\nu_{I_{\rm c}}}(t_{I_{\rm c, p}}))}{ln(\nu_{X}/\nu_{I_{\rm c}})+1.5ln(t_{X}/t_{I_{\rm c, p}})} \notag\\
\end{align}

Using the observed X-ray flux $F_{\nu_X}(1.19 \, \rm days) = (5.67\pm 1.44\times 10^{-2}) \, \rm \mu Jy$ at $1 \, \rm keV$  \citep{Yangyuhan2023} and radio flux $F_{\nu_r}(10.69 \, \rm days) = 92 \pm 22 \, \rm \mu Jy$ at $5.5 \, \rm GHz$  \citep{2023JWST}, we  derive  $p=2.52\pm 0.05$ and $\nu_{\rm m}(t_{I_{\rm c, p}})=(3.29\pm 1.62)\times 10^{14} \, \rm Hz$. The value of $\nu_{\rm m}(t_{I_{\rm c, p}})$ is in good agreement with the frequency of the $I_{\rm c}$ band ($2.69\times 10^{14} \, {\rm Hz} < \nu < 5.17\times 10^{14} \, {\rm Hz}$), implying that the peak is indeed caused by $\nu_{\rm m}$ crossing the optical band. Also, the electron spectrum index $p=2.52\pm 0.05$ is consistent with the observed X-ray photon index $\Gamma_{X}=1.73\pm0.10$ measured by XMM-Newton \citep{2023ApJ...956...97M}.

\bibliography{sample631}{}
\bibliographystyle{aasjournal}

\end{document}